# *Estimating Cost Savings from Early Cancer Diagnosis*

Zura Kakushadze,[§†1] Rakesh Raghubanshi[‡2] and Willie Yu[¶3]

[§] *Quantigic® Solutions LLC,[4] 1127 High Ridge Road, #135, Stamford, CT 06905*

[†] *Free University of Tbilisi, Business School & School of Physics
240, David Agmashenebeli Alley, Tbilisi, 0159, Georgia*

[‡] *Two29 Consulting LLC, 46 Sewell Avenue, Piscataway, NJ 08854*

[¶] *Centre for Computational Biology, Duke-NUS Medical School
8 College Road, Singapore 169857*

June 4, 2017; revised: August 23, 2017

Abstract

We estimate treatment cost-savings from early cancer diagnosis. For breast, lung, prostate and colorectal cancers and melanoma, which account for more than 50% of new incidences projected in 2017, we combine published cancer treatment cost estimates by stage with incidence rates by stage at diagnosis. We extrapolate to other cancer sites by using estimated national expenditures and incidence rates. A rough estimate for the U.S. national annual treatment cost-savings from early cancer diagnosis is in 11 digits. Using this estimate and cost-neutrality, we also estimate a rough upper bound on the cost of a routine early cancer screening test.

---

[1] Zura Kakushadze, Ph.D., is the President and a Co-Founder of Quantigic® Solutions LLC and a Full Professor in the Business School and the School of Physics at Free University of Tbilisi. Email: zura@quantigic.com

[2] Rakesh Raghubanshi, B.A. is an independent consultant with over 20 years of experience in the field of pharmaceutical research and development and has consulted for such companies as Cordis Corporation (then a Johnson & Johnson company, now a Cardinal Health company), Bayer AG and Ferring Pharmaceuticals. Email: rakesh.raghubanshi@gmail.com

[3] Willie Yu, Ph.D., is a Research Fellow at Duke-NUS Medical School. Email: willie.yu@duke-nus.edu.sg

## 1. Introduction

According to the Centers for Medicare & Medicaid Services (CMS), in 2015 the U.S. national health expenditure (NHE) was $3.2 trillion and accounted for 17.8% of Gross Domestic Product (GDP); NHE is projected to grow at an average rate of 5.6% per year in 2016-2025 [CMS, 2017]. Cancer care is projected to account for up to $177 billion in 2017 [NCI, 2017b], or nearly 1% of GDP.[5] American Cancer Society estimates 1.7 million new cases of cancer and 600 thousand deaths in 2017 [ACS, 2017]. Although overall incidence rates for new cancer cases have been falling on average 1.1% each year over the last 10 years and death rates have been falling on average 1.5% each year over 2005-2014 [NCI, 2017d], the impact of population changes in the U.S. on cancer prevalence may exceed the impact of declining cancer incidence rates. Also, the population in the U.S. is expected to become much older: by 2030, more than 20% of the U.S. residents are projected to be aged 65 and over, compared with 13% in 2010 [Ortman *et al*, 2014]. Since cancer incidence typically is higher in the elderly, the aging population and costly advancements in treatment options will impact cancer survival and care expenditures, both of which are likely to increase in the future. Overall, the cancer costs do tend to rise [NCI, 2017b].

Detecting and treating cancer at an early stage can and does save lives. Survival rates improve dramatically when cancer is diagnosed early and the disease is confined to the organ of origin, before it has had a chance to spread and is more likely to be treated successfully [Cho *et al*, 2014], [Aravanis *et al*, 2017]. Conversely, a late stage diagnosis essentially means that the cancer has spread making treatment much more difficult, thereby reducing chances of survival. Thus, according to [Cancer Research UK, 2017]: more than 9 in 10 bowel cancer patients will survive the disease for more than 5 years if diagnosed at the earliest stage; more than 90% of women diagnosed with breast cancer at the earliest stage survive their disease for at least 5 years compared to around 15% for women diagnosed with the most advanced stage of disease; more than 90% of women diagnosed with the earliest stage ovarian cancer survive their disease for at least 5 years compared to around 5% for women diagnosed with the most advanced stage of disease; around 70% of lung cancer patients will survive for at least a year if diagnosed at the earliest stage compared to around 14% for people diagnosed with the most advanced stage of disease. The importance of diagnosing cancer early for survival cannot be overstated.

Another important aspect relating to early diagnosis is treatment costs. Thus, cancer patient costs of care in the last year of life are sizably higher than during earlier stages [NCI, 2017c]. Further, in many cases, it is much less costly to treat cancer when it is diagnosed early.[6]

---

[5] The $177B figure is a high estimate assuming incidence/survival rate trends and 5% cost increases [NCI, 2017b].

[6] E.g., for later-stage melanoma, chemotherapy, etc., sizably increase costs [Styperek and Kimball, 2012] (see Table V therein).

Considering that the wealth of nations is not limitless, one of our better chances to reduce staggering cancer treatment costs is through early detection and intervention. Traditionally, cancer research has focused on treatments for late-stage disease, encompassing an estimated 85% of the annual allocation [Curry *et al*, 2003]. Thus, global oncology drug costs reached $107 billion in 2015 and are projected to exceed $150 billion by 2020 [Buffery, 2016]. Such trends appear to be producing a shift in thinking amongst various stakeholders, such as policy makers, payers, providers, and consumers, in reorienting research toward prevention[7] and early detection. Recent high fund-raising figures by companies such as Grail, Inc., which raised close to $1B in its recent series B funding round [Nasdaq GlobeNewswire, 2017], and Guardant Health, which recently raised $360M from investors (bringing its total raised to $550M) [Herper, 2017] speak volumes in this regard. Therefore, here we ask the following question:

*What are the estimated cost-savings from early cancer diagnosis?* Our goal is to arrive at a *conservative* estimate. Therefore, we define cost-savings from early diagnosis by assuming that all stage III and IV cases are detected at stages I or II, with the current incidence rates therefor. We specifically exclude stage 0. The requisite data is scarce, incomplete and scattered. We use costs and incidence rates data available for 4 and 19 cancer sites, respectively, and extrapolate to various other cancers. We focus on the U.S. expenditures. While healthcare costs in other countries are in many cases lower than in the U.S., the estimates apply directionally worldwide. Finally, the cost-savings estimates hereof are limited to direct costs for treatment only. When conducting health economic analyses, a critical piece of the evaluation is the question "what is the value of health, both to the individual patient and to the overall system as a whole?" In considering this question, indirect financial costs of cancer treatment can be an additional burden to the people diagnosed with cancer, their families, their employers, and the society in general, and the added costs can be significant. However, as mentioned above, we are after a conservative estimate and such considerations would only add to it. Our estimate, $26B/year, is by no means "precise" as it is extrapolated. However, it is likely correct within a factor of 2.

The remainder is organized as follows. Section 2 discusses i) a methodology for estimating cost-savings from early cancer diagnosis and ii) data based on commercial claims for breast cancer as reported in [Blumen *et al*, 2016] and on incidence rates by stage at diagnosis as reported in [Iqbal *et al*, 2015]. Section 3 discusses incidence rate data for 19 cancer sites as reported in [Morris *et al*, 2013] (and also [Parikh-Patel *et al*, 2015]). Section 4 discusses Medicare claims data as reported in [Schrag, 2015] for 4 cancer sites. Section 5 discusses melanoma data as reported in [Styperek and Kimball, 2012]. Section 6 discusses i) extrapolation to other cancer sites and ii) national expenditure estimates as reported in [NCI, 2017b]. Section 7 briefly concludes. Tables 1 through 11 contain data utilized in our analysis.

---

[7] Reducing exposure to known carcinogens [Loeb and Harris, 2008], [Ananthaswamy and Pierceall, 1990] such as tobacco, etc., is important. However, cancer occurs at the DNA level via somatic mutations (see [Goodman and Fygenson, 1998], [Lindahl, 1993], [Tomasetti *et al*, 2017] and referenced therein) and is not always preventable.

## 2. Breast Cancer

### *2.1. Costs by Stage at Diagnosis: Commercially Insured Population Study*

Blumen *et al* [2016] analyze commercially insured U.S. women aged 18 to 64 years who were newly diagnosed with breast cancer in 2010.[8] Table 1, which is adapted from [Blumen *et al*, 2016], summarizes their results. The average per-patient allowed costs in the 12 months following diagnosis are $60,637, $82,121, $129,387 and $134,682 for stages 0, I/II, III, and IV at diagnosis, respectively. The average per-patient allowed costs in the 24 months following diagnosis are $71,909, $97,066, $159,442 and $182,655 for stages 0, I/II, III, and IV at diagnosis, respectively. These costs are not adjusted for inflation or any other temporal changes.

### *2.2. Incidence Rates by Stage at Diagnosis*

Iqbal *et al* [2015] analyze various data for 452,215 women diagnosed with invasive breast cancer from 2004 to 2011 who were identified in the Surveillance, Epidemiology, and End Results (SEER) 18 registries database.[9] This study focuses on stages I, II, III and IV, specifically excluding in situ stage 0 and stage unknown cases.[10] Table 2, which is adapted from [Iqbal *et al*, 2015], summarizes the data for breast cancer incidence rates (in %) by stage (I, II, III and IV only) at diagnosis, including aggregated numbers as well as those broken down by eight racial/ethnic groups, across which there is some degree of variability, which should be kept in mind when interpreting cost-savings estimates. We will use the figures (column 2, Table 2) aggregated across all racial/ethnic groups: 48%, 34.6%, 12.4% and 5% for stages I, II, III and IV, respectively.

---

[8] This study utilizes the Truven Health MarketScan® commercial claims database using 2010 as the index year, 2009 as a look-back year, and 2011 and 2012 as the 24-month look-forward period. It infers the stage – to wit, stage 0, I/II, III, or IV – at diagnosis based on identification of stage-specific treatments recommended in the National Comprehensive Cancer Network (NCCN) treatment guidelines [NCCN, 2017]. Cases at stages I and II are combined as the NCCN treatment recommendations are interchangeable for these stages. See [Blumen *et al*, 2016] for details.

[9] According to [NCI, 2017a]: the SEER 18 registries consist of the SEER 13 plus Greater California, Greater Georgia, Kentucky, Louisiana, and New Jersey; the SEER 13 registries consist of the SEER 9 plus Los Angeles, San Jose-Monterey, Rural Georgia and the Alaska Native Tumor Registry; the SEER 9 registries are Atlanta, Connecticut, Detroit, Hawaii, Iowa, New Mexico, San Francisco-Oakland, Seattle-Puget Sound, and Utah. The SEER 18 covers about 28% of the total U.S. population (based on the 2010 Census) [Iqbal *et al*, 2015].

[10] It also excludes a small fraction of cases with borderline, undetermined or unknown estrogen receptor status, and those with prior history of any cancer, leaving 373,563 cases in the study [Iqbal *et al*, 2015].

## 2.3. Cost-Savings Estimates

To estimate cost-savings of early (stages I and II) vs. late (stages III and IV) diagnoses, we use the 12-month and 24-month average per-patient allowed costs for stages I/II, III and IV from Blumen *et al* [2016] (see Subsection 2.1 and columns 4 and 6 of Table 1) and incidence rates by stage at diagnosis (see Subsection 2.2 and column 2 in Table 2) from [Iqbal *et al*, 2015]. The average 12- and 24-month estimated per-patient cost-savings from early diagnosis are given by

$$S_{12-mo} = (X_{III} - X_{I,II}) \times R_{III} + (X_{IV} - X_{I,II}) \times R_{IV} \tag{1}$$

$$S_{24-mo} = (Y_{III} - Y_{I,II}) \times R_{III} + (Y_{IV} - Y_{I,II}) \times R_{IV} \tag{2}$$

Here $X_{I,II} = \$82,121$, $X_{III} = \$129,387$ and $X_{IV} = \$134,682$ are the 12-month average per-patient allowed costs for stages I/II, III and IV, respectively; $Y_{I,II} = \$97,066$, $Y_{III} = \$159,442$ and $Y_{IV} = \$182,655$ are the 24-month average per-patient allowed costs for stages I/II, III and IV, respectively. The incidence rates by stage at diagnosis are $R_I = 48.0\%$, $R_{II} = 34.6\%$, $R_{III} = 12.4\%$ and $R_{IV} = 5.0\%$ for stages I, II, III and IV, respectively. Thus, in Eq. (1) and Eq. (2) we are estimating average savings by assuming that all stage III and IV cases are diagnosed early, at stage I or II. With these assumptions, the estimated cost-savings are $S_{12-mo} = \$8,489$ and $S_{24-mo} = \$12,014$ (these figures are rounded to the nearest integer). It is also instructive to estimate relative (as opposed to absolute) cost-savings compared with average per-patient costs across all stages. The latter can be estimated as

$$C_{12-mo} = X_{I,II} \times (R_I + R_{II}) + X_{III} \times R_{III} + X_{IV} \times R_{IV} \tag{3}$$

$$C_{24-mo} = Y_{I,II} \times (R_I + R_{II}) + Y_{III} \times R_{III} + Y_{IV} \times R_{IV} \tag{4}$$

These estimates, $C_{12-mo} = \$90,610$ and $C_{24-mo} = \$109,080$, are based on stage I, II, III and IV cases only. However, if we include in situ stage 0 cases, then the average per-patient costs are lower. To estimate these costs, we need the incidence rate for stage 0 cases. Thus, according to [Siegel *et al*, 2017], the estimated number of new in situ (stage 0) female breast cancer cases in the U.S. in 2017 is 63,410, whereas the estimated number of new invasive (stages I, II, III and IV) female breast cancer cases in the U.S. in 2017 is 252,710. We will take $\tilde{R}_0 = 63,410 / (252,710 + 63,410) = 20.1\%$ as the incidence rate for stage 0. Accordingly, the incidence rates for stage I-IV cases are given by $\tilde{R}_K = (1 - \tilde{R}_0) \times R_K$ with $K = I, II, III, IV$. Note that $R_I + R_{II} + R_{III} + R_{IV} = 100\%$, and, therefore, $\tilde{R}_0 + \tilde{R}_I + \tilde{R}_{II} + \tilde{R}_{III} + \tilde{R}_{IV} = 100\%$.

The average per-patient costs across all stages including stage 0 then are given by

$$\tilde{C}_{12-mo} = X_0 \times \tilde{R}_0 + X_{I,II} \times (\tilde{R}_I + \tilde{R}_{II}) + X_{III} \times \tilde{R}_{III} + X_{IV} \times \tilde{R}_{IV} \tag{5}$$

$$\tilde{C}_{24-mo} = Y_0 \times \tilde{R}_0 + Y_{I,II} \times (\tilde{R}_I + \tilde{R}_{II}) + Y_{III} \times \tilde{R}_{III} + Y_{IV} \times \tilde{R}_{IV} \qquad (6)$$

Here $X_0 = \$60,637$ and $Y_0 = \$71,909$ (see Subsection 2.1), so $\tilde{C}_{12-mo} = \$84,598$ and $\tilde{C}_{24-mo} = \$101,624$, which are 6.6% and 6.8% lower than $C_{12-mo}$ and $C_{24-mo}$, respectively.

The relative cost-savings are defined as

$$E_{12-mo} = S_{12-mo} / C_{12-mo} \qquad (7)$$

$$E_{24-mo} = S_{24-mo} / C_{24-mo} \qquad (8)$$

$$\tilde{E}_{12-mo} = \tilde{S}_{12-mo} / \tilde{C}_{12-mo} = (1 - \tilde{R}_0) \times S_{12-mo} / \tilde{C}_{12-mo} \qquad (9)$$

$$\tilde{E}_{24-mo} = \tilde{S}_{24-mo} / \tilde{C}_{24-mo} = (1 - \tilde{R}_0) \times S_{24-mo} / \tilde{C}_{24-mo} \qquad (10)$$

These estimated relative cost-savings are $E_{12-mo} = 9.37\%$, $E_{24-mo} = 11.01\%$, $\tilde{E}_{12-mo} = 8.02\%$, $\tilde{E}_{24-mo} = 9.45\%$. So, *roughly*, we expect around 8-11% savings from early diagnosis. When we include stage 0, in Eq. (9) and Eq. (10) the average 12- and 24-month estimated per-patient cost-savings $\tilde{S}_{12-mo}$ and $\tilde{S}_{24-mo}$ are computed via Eq. (1) and Eq. (2), respectively, but with $R_{III}$ and $R_{IV}$ replaced by $\tilde{R}_{III}$ and $\tilde{R}_{IV}$, hence the factor $(1 - \tilde{R}_0)$. Therefore, including stage 0 reduces the absolute costs-savings and also to a lesser degree the average costs, so overall the relative cost-savings are also reduced. Thus, we have $\tilde{E}_{12-mo} / E_{12-mo} = 85.59\%$, and $\tilde{E}_{24-mo} / E_{24-mo} = 85.83\%$, so including stage 0 reduces the relative cost-savings by about 14-15%. Generally, stage unknown can also alter these figures, but to a lesser degree.

## 3. Incidence Rates by Stage at Diagnosis: 19 Cancers

Morris *et al* [2013] provide detailed data for stage at diagnosis for California adults aged 20 and older diagnosed with cancer during 2005-2009. Their data contains 19 cancer sites. We compile their data into Table 3, which provides (for each of the 19 cancer sites) the total number of cases, numbers of cases for stages 0-IV and stage unknown, and the corresponding incidence rates (in %), both including and excluding stage 0 and stage unknown. For some cancer sites stage 0 data is not available (NA). For bladder stage 0 and stage I are combined.

Comparing incidence rates in columns 15-18 of Table 3 (these correspond to stages I-IV only, with stage 0 and stage unknown excluded) for breast cancer, we see that they are very close to the incidence rates in column 2 of Table 2 obtained from [Iqbal *et al*, 2015], which is based on the SEER 18 database (see Subsection 2.2) and includes various other U.S. regions.

Morris *et al* [2013] also provide data for various racial/ethnic and age groups. We compile their data for breast cancer into Table 4. The last 4 rows correspond to the incidence rates by stage at diagnosis with stage 0 and stage unknown excluded. The racial/ethnic group incidence rates are consistent with those in Table 2, which are based on the SEER 18 [Iqbal *et al*, 2015].

### *3.1. A Sanity Check: 5 Cancers*

Parikh-Patel *et al* [2015] analyze data by stage at diagnosis, quality of treatment, and survival among persons diagnosed with breast, colon, rectal, lung, and prostate cancer in California between 2004 and 2012. We compile their data into Table 5. The last 4 columns of Table 5 are consistent with those in Table 3. A notable difference exists for prostate cancer, for which there are relatively few cases at stage I in Table 3, and a more sizable number in Table 5, but the stage I + II incidence rates from Table 3 (85.5%) and Table 5 (84.64%) are still consistent.

## 4. Medicare Data: 4 Cancers

Schrag [2015] provides Medicare spending estimates for breast, colorectal, lung and prostate cancers in California between 2007-2011. One of the reasons cited in [Schrag, 2015] for focusing on Medicare spending is that "Medicare data, unlike data from other payers, are readily accessible". We compile the data from [Schrag, 2015] into Table 6 (mean per-patient Medicare spending in the first year after diagnosis by stage at diagnosis) and Table 7 (mean per-patient Medicare spending in the last year of life by stage at diagnosis). The costs in Table 7 in the last year of life are relatively uniform with the stage at diagnosis. However, the costs in Table 6 in the first year after diagnosis increase considerably with the stage at diagnosis.

For the first year after diagnosis, we can use the same methodology as in Subsection 2.3 to estimate cost-savings from early (stages I and II) vs. late (stages III and IV) diagnosis. Here we can estimate the following quantities: $C_{12-mo}$ (average per-patient cost in the first 12 months after diagnosis based on stages I-IV) using Eq. (3); $S_{12-mo}$ (average per-patient cost-savings from early diagnosis in the first 12 months from diagnosis) using Eq. (1); and $E_{12-mo}$ (the relative cost-savings) using Eq. (7). In Eq. (1) and Eq. (3) the quantity (the stage I/II cost)

$$X_{I,II} = (X_I \times R_I + X_{II} \times R_{II}) / (R_I + R_{II}) \qquad (11)$$

Here $X_K$ is the stage $K$ cost from Table 6 (columns 2-5), and $R_K$ is the stage $K$ fraction from Table 6 (columns 6-9) with $K = I, II, III, IV$. For prostate cancer we set $R_I = 0$, so $X_{I,II} = X_{II}$.

The results are summarized in Table 8. For breast cancer the relative savings (11.35%) are consistent with our estimates in Subsection 2.3 for commercially insured patients. Note that the fractions in columns 6-9 of Table 6 are somewhat different from all-age incidence rates in, e.g., Table 3. For instance, for breast cancer in Table 6 we have 52% for stage I, while in Table 3 we have about 48%. This difference appears to be due to the age group (Medicare). Thus, for age 65+ we have the incidence rate close to 54% for breast cancer in Table 4. Also, let us note that the average per-patient costs reported in [Schrag, 2015] in the graph at p.9 are somewhat lower[11] than those in column 2 of Table 8 (which are computed based on the data in Table 6 as set forth above). This difference may be due to stage 0 and/or stage unknown contributions. There is not enough information to determine this; however, the difference is small (< 7.52%).

Finally, let us mention that our estimate for the average 12-month cost for breast cancer for Medicare (see Table 8) is $38,130, while the analogous estimate for the commercially insured population from Subsection 2.3 is $90,610, i.e., the Medicare figure is about 42% of the commercial insurance figure.[12] However, this ratio is by no means "precise" as the commercial insurance figures in [Blumen *et al*, 2016] are from 2010 diagnoses, whereas the Medicare data in [Schrag, 2015] is from 2007-2011 diagnoses, so the actual ratio could be higher. However, the ballpark appears to be correct. Thus, according to Appendix D of [Pyenson *et al*, 2016],[13] in 2004 the cancer population was 63,935 (commercial), and 118,089 (Medicare), while the total spending (allowed) in this population was $2,281,981,711 (commercial) and $2,619,153,436 (Medicare), so the per-patient spending in 2004 was $35,692 (commercial) and $22,179 (Medicare), and the corresponding Medicare-to-commercial ratio was about 62%. According to this source, in 2014 the cancer population was 264,204 (commercial), and 133,225 (Medicare), while the total spending (allowed) in this population was $13,908,337,950 (commercial) and $3,672,799,298 (Medicare), so the per-patient spending in 2014 was $52,642 (commercial) and $27,568 (Medicare), and the corresponding Medicare-to-commercial ratio was about 52%. So, the rough ratio for breast cancer we obtained above from the [Blumen *et al*, 2016] and [Schrag, 2015] data is in the ballpark of those based on the [Pyenson *et al*, 2016] data (for all cancers).[14]

## 5. Melanoma

Styperek and Kimball [2012] provide malignant melanoma costs by stage at diagnosis based on 2008 data.[15] Based on [Styperek and Kimball, 2012], we compile the costs in the first year after diagnosis and incidence rates by stage at diagnosis in Table 9. As in the previous section, we can estimate the following quantities: $C_{12-mo}$ (average per-patient cost in the first 12 months after diagnosis based on stages I-IV) using Eq. (3); $S_{12-mo}$ (average per-patient cost-savings from early diagnosis in the first 12 months from diagnosis) using Eq. (1); and $E_{12-mo}$ (the relative cost-savings) using Eq. (7). In Eq. (1) and Eq. (3) the quantity (the stage I/II cost) $X_{I,II}$ is given by Eq. (11). Using the data for stages I-IV in Table 9, we have $C_{12-mo} = \$12{,}541$, $S_{12-mo} = \$5{,}085$ and $E_{12-mo} = 40.55\%$. Such a dramatic cost reduction from early diagnosis is due to more than a 3-fold increase in melanoma treatment costs between stages II and III. Using the melanoma incidence rates from Table 3, we would get a higher $E_{12-mo} = 41.41\%$.

---

[11] To wit, $35,264 for breast cancer, $28,213 for prostate cancer, $78,444 for lung cancer, and $69,687 for colorectal cancer. [Schrag, 2015] cites "CCR-Medicare, 2014 data linkage, Healthcare Delivery Research Program, National Cancer Institute" as the source for these average spending figures.

[12] The $35,264 figure from [Schrag, 2015] (see fn.11 hereof), divided by $\tilde{C}_{12-mo} = \$84,598$ (see Subsection 2.3) gives 41.7%.

[13] Based on Milliman analysis of the 2004-2014 Truven Health MarketScan® data and Medicare 5% sample data.

[14] The aforesaid Medicare (based on [Schrag, 2015]) and commercial (based on [Blumen *et al*, 2016]) data are not necessarily uniformly normalized. On the other hand, we expect that the Medicare and commercial figures from Appendix D of [Pyenson *et al*, 2016] are uniformly normalized and can be meaningfully compared with each other.

[15] There is a variability in costs reported in various studies [Guy Jr *et al*, 2012]. Also see [Alexandrescu, 2009].

## 6. Extrapolating to Other Cancers

### 6.1. Relative Cost-Savings

Using the definitions in Eq. (1), Eq. (3) and Eq. (7), we can rewrite the relative cost-savings $E_{12-mo}$ as follows:

$$E_{12-mo} = F / (1 + F) \tag{12}$$

$$F = R_{Late} \times G \tag{13}$$

Here $R_{Late} = R_{III,IV} = R_{III} + R_{IV}$ is the incidence rate of late-stage (stages III and IV) diagnosis, and

$$G = X_{Late} / X_{Early} - 1 \tag{14}$$

We define $X_{Late} = X_{III,IV} = (X_{III} \times R_{III} + X_{IV} \times R_{IV}) / R_{Late}$, and $X_{Early} = X_{I,II}$ is defined in Eq. (11). Therefore, the relative-cost savings are controlled by two parameters, $R_{Late}$ and $G$.

We summarize these quantities in Table 10 for breast cancer (based on the commercial claims data from [Blumen *et al*, 2016] and the Medicare data from [Schrag, 2015]), prostate, lung and colorectal cancers (based on the Medicare data from [Schrag, 2015]), and melanoma (based on the data from [Styperek and Kimball, 2012]). There is a substantial heterogeneity in both $R_{Late}$ and $G$, including in commercial vs. Medicare data. Lung cancer has higher $E_{12-mo} = 25.08\%$ largely due to it mostly being diagnosed late ($R_{Late} = 74\%$). On the other hand, melanoma has really high $E_{12-mo} = 40.55\%$ mainly due to a large jump in the treatment costs between early ($X_{Early} = \$7{,}456$) and late ($X_{Late} = \$40{,}641$) diagnoses. In contrast, for prostate cancer we have low $E_{12-mo} = 6.26\%$ due low $R_{Late} = 16\%$ as well as low $G = 0.42$.

Table 3 contains incidence rates by stage at diagnosis for 19 cancer sites. We will use this data for cancer sites (for which cost data is available – see below)[16] beyond the 5 cancers we discuss above. For such cancer sites we will use the mean value of $R_{Late}$ based on the last 2 columns of Table 3. This mean value is $\bar{R}_{Late} = 39.94\%$ with a standard deviation[17] = 21.04%.

---

[16] E.g., cost data are available for colorectal cancer, not for colon cancer or rectum carcinoma separately; etc.

To estimate $F$ via Eq. (13), we also need the values of $G$. For breast, prostate, lung and colorectal cancers and melanoma, we will take these values from Table 10. Conservatively, for breast cancer we will take the lower value from row 2 of Table 10. For cancer sites beyond these 5, we must estimate $G$. We exclude row 3 (the higher value of $G$ for breast cancer) and row 7 (melanoma, which is an outlier). The remaining 4 values of $G$ (rows 2 and 4-6 in Table 10) have the mean of 0.5202 and the median of 0.5232. We will set $G = 0.5202$ for the other sites.

### 6.2. Estimated National Expenditures

Estimated national expenditures between 2010 and 2020 are provided in [NCI, 2017b][18] for 17 cancer sites (which are not the same as those in Table 3). Detailed data can be downloaded from the webpage referenced in [NCI, 2017b]. For each year, including 2017 which we focus on, this data contains 6 numbers for the estimated national expenditures based on 6 different assumptions, to wit: i) both incidence and survival are constant; ii) incidence follows recent trends, survival is constant; iii) survival follows recent trends, incidence is constant; iv) both incidence and survival follow recent trends; v) both incidence and survival follow recent trends, annual costs increase at 2% (applied to initial and last phases);[19] vi) both incidence and survival follow recent trends, annual costs increase at 5% (applied to initial and last phases). In i)-iv) above annual costs are assumed to be constant. For estimated national costs for 2017 we take the mean of these six figures, and the so-averaged figures are in column 2 of Table 11, while the corresponding standard deviations (in %) are in column 3 thereof. These standard deviations are reasonably small. For the cancer sites in column 1 of Table 11 we also give the expected 2017 new incidence rates (as reported in [ACS, 2017] and [Siegel *et al*, 2017]) in column 4. Dividing column 2 by column 4 produces column 5, the per-new-incidence estimated costs.[20] Columns 6-8 list the values of $G$, $R_{Late}$ and $E_{12-mo}$ as set forth above (also see the caption to Table 11). We then *roughly* estimate national cost-savings $S_{nat}$ from early diagnosis via

$$S_{nat} = C_{nat} \times E_{12-mo} \times H \tag{15}$$

$$H = \tilde{E}_{12-mo} / E_{12-mo} = 1 / (1 + X_0 \tilde{R}_0 / C_{12-mo} (1 - \tilde{R}_0)) \tag{16}$$

---

[17] For the same data in Table 3, median = 42.4% and MAD = 28.9% (MAD = mean absolute deviation). Below we will use the lower mean value $\bar{R}_{Late} = 39.94\%$ and not the higher median value for our conservative estimate.

[18] Also see [NCI, 2017c].

[19] Initial phase = initial year after diagnosis; last phase = last year of life; continuing phase = in-between period. For detailed information about the methods, data sources and assumptions in [NCI, 2017b], see [Mariotto, 2011].

[20] Which are not the same as the per-patient costs. We give column 5 in Table 11 for orientation purposes.

Here the factor $H$ corrects for the stage 0 contributions (see Subsection 2.3). For breast cancer we take $H = 85.59\%$ (see Subsection 2.3). For melanoma, using $\tilde{R}_0 = 40.08\%$ from Table 3, $X_0 = \$984$ from Table 9, and $C_{12-mo} = \$12,541$ from Section 5, we get $H = 95.01\%$, so the reduction due to stage 0 is small despite a large $\tilde{R}_0$ as the stage 0 cost $X_0$ is small. For other cancer sites $\tilde{R}_0$ is sub-10% (or NA) and assuming that $X_0$ is sizably smaller than $C_{12-mo}$ (see Subsection 2.3), for these cancer sites $H$ is expected to be close to 1. The results for estimated national costs-savings $S_{nat}$ are in column 9 of Table 11 (and the national cost estimates $C_{nat}$ are taken from column 2 thereof). The factor $H$ is set to 85.59% for breast cancer, and to 1 for other cancer sites. The all-sites national cost-savings estimate is $26B.

*6.3. Caveats*

Our estimates are clearly far from being "precise" for a variety of reasons, including extrapolating $G$ and $R_{Late}$ to various cancer sites based on data available for 4 and 19 cancer sites, respectively. However, we have taken a conservative approach to so-extrapolating $G$. Nonetheless, e.g., for certain cancers cost-savings from early diagnosis may be less attainable than from others.[21] Also, in Eq. (15) we simply use $E_{12-mo}$ (estimated relative cost-savings for the first year after diagnosis).[22] The last phase (the last year of life) costs are skewed toward higher figures (see, e.g., [NCI, 2017c]) due to added expenses at this phase. However, with early diagnosis the survival rate would also go up thereby decreasing the contribution of the last phase to the overall costs. Another caveat is that commercial insurance vs. Medicare vs. other payer costs are heterogeneous (and the corresponding data is not readily available), the costs have strong temporal dependence as new treatments become available continuously (and most available data is at least some years out-of-date), etc., so cost-estimates are uneven (see, e.g., [Guy Jr. *et al*, 2012]). However, the $26B figure above is likely correct within a factor of 2.

## 7. Conclusions

The above rough estimate for cost-savings from early cancer diagnosis, $26B, is only about 17% of the total estimated expenditures (see Table 11) and appears to be reasonable despite various built-in (conservative) extrapolations. If we take breast, lung, prostate and colorectal cancers and melanoma, which are top 5 cancers by incidence with the total 859,110 estimated new cases in 2017, which amounts to 50.87% of all 1,688,780 estimated new cases across all

---

[21] Let us note a minor caveat that for bladder cancer the stage 0 and stage I figures in Table 3 are combined.

[22] Albeit adjusted for stage 0 contributions via the factor $H$ (see above). Also, note that the rates $E_{12-mo}$ and $E_{24-mo}$ are consistent with each other (see the discussion after Eq. (10) in Subsection 2.3), so using the $E_{12-mo}$ rates in our extrapolated estimations is reasonable. A more important caveat is related to the last year of life costs, which are skewed, and which we discuss below in this Subsection.

cancer sites,[23] the estimated costs add up to over $67B (or about 43.87% of costs for all sites), and the corresponding estimated cost-savings from early diagnosis add up to over $10.7B (or about 41.49% of cost-savings for all sites). Again, these figures should be considered keeping in mind the caveats we discuss above. Even assuming only 50%, the cost-savings are staggering.[24]

Above we focus on the U.S. Cost-savings from early cancer diagnosis in some other regions of the world have been addressed in the literature. Here we will not attempt an exhaustive review. Instead, keeping in mind that generally healthcare costs in other countries are in many cases lower than in the U.S., let us mention a U.K. study [Incisive Health, 2014][25] and a shorter summary thereof [Birtwistle, 2014], according to which the fractions of the costs for stage I vs. stage IV at diagnosis for colon, rectal, ovarian and lung cancers are approximately 27.2%, 37.3%, 35.1% and 61.1%, respectively. Again, cost-savings from early diagnosis are staggering.

It is precisely these economic considerations that underlie recent high fund-raising figures by companies such as Grail, Inc., which raised close to $1B in its recent series B funding round [Nasdaq GlobeNewswire, 2017], and Guardant Health, which recently raised $360M from investors (bringing its total raised to $550M) [Herper, 2017]. These figures may appear inflated at first, but are not unreasonable based on the estimated annual cost-savings we discuss here. Early cancer diagnosis does not only save lives but will also save billions of dollars in costs.

In this regard, we can estimate a rough upper bound on the cost of a routine early cancer screening test. We have estimated $26B/yr in savings from early cancer diagnosis. According to [Mehrotra *et al*, 2007], there were about 44.4M adults annually who received preventive health examinations during 2002-2004. Let us take this figure as a rough estimate for the number of annual early cancer screening tests. Then on a cost-neutral basis an approximate upper bound for the cost of such a test is $600. Let us note that this estimate could actually be lower or higher depending on various details. First, our $26B/yr estimate is conservative and the actual cost-savings could be higher. Second, this estimate is extrapolated to all cancer sites. Early screening tests may apply to a limited number of cancer sites, and the available cost-savings would be lower. However, if so, then the number of patients screened may also be limited to those at risk, which would decrease the number of tests administered. Third, as

---

[23] These figures relate to invasive cancer incidences. Thus, in addition, e.g., about 63,410 cases of female breast cancer in situ and 74,680 cases of melanoma in situ are expected to be diagnosed in 2017 [Siegel *et al*, 2017].

[24] Also, as mentioned above, here we are not including indirect costs of cancer or considerations stemming from the quality-adjusted life-years (QALY), etc. Again, our goal here is to arrive at a reasonable conservative estimate.

[25] Also, see, e.g., [Laudicella *et al*, 2016].

mentioned above, we are not including indirect costs of cancer or considerations stemming from the quality-adjusted life-years (QALY), etc., which may also increase said upper bound.[26]

Let us emphasize that, here our goal is *not* to determine cost-savings from any particular early cancer screening (be it blood-based or any other such) test. Realistically, all such tests are expected to have false negatives and false positives. Clearly, false negatives would decrease any cost-savings associated with early cancer detection. However, if the rate of false negatives is too high to begin with, such a test may not be viable in the first instance. On the other hand, false positives could increase costs as they may cause unnecessary additional testing and/or treatment, not to mention all the anxiety and stress to the patients misdiagnosed with cancer by such false positives. Again, if the rate of false positives is too high to begin with, such a test may not be viable in the first instance. Without specific and reliable data (which does not exist yet) from, say, blood tests, it is virtually impossible to intelligently estimate the effects of false positives or negatives and such an estimate would at best be highly speculative and likely uninformative. Thus, currently it is unknown what the rates of false positives or negatives will be for new cancer screening technologies such as ctDNA (circulating tumor DNA) based blood tests – these technologies are still in nascent stages [Aravanis *et al*, 2017]. Instead, our goal here is to *conservatively estimate* the size of cost-savings from early detection (which is roughly the size of the "market", which affects the pricing of early detection tests as discussed above). Our estimate is only rough for the multitude of reasons discussed in detail above and our $26B/yr figure likely is accurate within a factor of 2. However, this figure is reasonable and there is value in knowing the order of magnitude of the available cost-savings. Thus, from our analysis it is clear that these cost-savings should be much larger than, say, $1-2B/yr, but at the same time they are unlikely to lead to 50% overall cost reduction (this, among other things, is due to high-incidence-level cancers such as prostate and breast cancers already being diagnosed early in many cases based on currently available screenings such as mammograms and prostate exams). However, by piecing together scattered (and not-so-readily available) data and being conservative, our estimates appear to be reasonable and in line with others (see fn.26).

## Acknowledgments

ZK would like to thank Bert Vogelstein for an email correspondence and a valuable suggestion to include the discussion on the costs of early cancer screening tests in the manuscript. We would also like to thank anonymous referees for suggestions, which improved the manuscript.

---

[26] Our aforesaid estimate $600 per test is consistent with Grail, Inc.'s projections of $1,000 per test, as reported in [Regalado, 2017].

# References

ACS (2017) *Cancer Facts & Figures 2017.* Atlanta, GA: American Cancer Society. Available online: https://www.cancer.org/content/dam/cancer-org/research/cancer-facts-and-statistics/annual-cancer-facts-and-figures/2017/cancer-facts-and-figures-2017.pdf.

Alexandrescu, D.T. (2009) Melanoma costs: A dynamic model comparing estimated overall costs of various clinical stages. *Dermatology Online Journal* **15**(11): 1. Available online: http://escholarship.org/uc/item/53f8q915.

Ananthaswamy, H.N.; Pierceall, W.E. (1990) Molecular mechanisms of ultraviolet radiation carcinogenesis. *Photochemistry and Photobiology* **52**(6): 1119-1136. Available online: http://onlinelibrary.wiley.com/doi/10.1111/j.1751-1097.1990.tb08452.x/epdf.

Aravanis, A.M.; Lee, M.; Klausner, R.D. (2017) Next-Generation Sequencing of Circulating Tumor DNA for Early Cancer Detection. *Cell* **168**(4): 571-574. Available online: http://www.cell.com/cell/pdf/S0092-8674(17)30115-0.pdf.

Birtwistle, M. (2014) Saving lives and averting costs? The case for earlier diagnosis just got stronger. London, UK: Cancer Research UK, September 2014. Available online: http://scienceblog.cancerresearchuk.org/2014/09/22/saving-lives-and-averting-costs-the-case-for-earlier-diagnosis-just-got-stronger/.

Blumen, H., Fitch, K., Polkus, V. (2016) Comparison of Treatment Costs for Breast Cancer, by Tumor Stage and Type of Service. *American Health & Drug Benefits* **9**(1): 23-32. Available online: https://www.ncbi.nlm.nih.gov/pmc/articles/PMC4822976/pdf/ahdb-09-023.pdf.

Buffery, D. (2016) Innovation Tops Current Trends in the 2016 Oncology Drug Pipeline. *American Health & Drug Benefits* **9**(4): 233-238. Available online: https://www.ncbi.nlm.nih.gov/pmc/articles/PMC5004820/pdf/ahdb-09-233.pdf

Cancer Research UK (2017) Why is early diagnosis important? London, UK: Cancer Research UK. Available online: http://www.cancerresearchuk.org/about-cancer/cancer-symptoms/why-is-early-diagnosis-important.

Cho, H.; Mariotto, A.B.; Schwartz, L.M.; Luo, J.; Woloshin, S. (2014) When do changes in cancer survival mean progress? The insight from population incidence and mortality. *Journal of the National Cancer Institute Monographs* **2014**(49): 187-197. Available online: https://www.ncbi.nlm.nih.gov/pmc/articles/PMC4841163/pdf/lgu014.pdf.

CMS (2017) *National Health Expenditure Data: NHE Factsheet.* Baltimore, MD: U.S. Centers for Medicare & Medicaid Services. Available online https://www.cms.gov/research-statistics-data-and-systems/statistics-trends-and-reports/nationalhealthexpenddata/nhe-fact-sheet.html.

Curry, S.J.; Byers, T.; Hewitt, M. (eds.) (2003) Fulfilling the Potential for Cancer Prevention and Early Detection. Washington, DC: National Academies Press, p. 371. Available online: http://www.nap.edu/catalog/10263.html.

Edge, S.B.; Byrd, D.R.; Compton, C.C.; Fritz, A.G.; Greene, F.L.; Trotti, III, A. (eds.) (2011) *American Joint Committee on Cancer (AJCC), Cancer Staging Manual.* New York, NY: Springer.

Goodman, M.F.; Fygenson, K.D. (1998) DNA polymerase fidelity: from genetics toward a biochemical understanding. *Genetics* **148**(4): 1475-1482. Available online: https://www.ncbi.nlm.nih.gov/pmc/articles/PMC1460091/pdf/9560367.pdf

Guy Jr, G.P.; Ekwueme, D.U.; Tangka, F.K.; Richardson, L.C. (2012) Melanoma Treatment Costs: A Systematic Review of the Literature, 1990–2011. *American Journal of Preventive Medicine* **43**(5): 537-545. Available online: https://www.ncbi.nlm.nih.gov/pmc/articles/PMC4495902/pdf/nihms704314.pdf.

Herper, M. (2017) *Guardant Health Raises $360 Million In Race To Create Cancer Blood Tests.* New York, NY: Forbes, May 2017. Available online: https://www.forbes.com/sites/matthewherper/2017/05/11/guardant-health-raises-350-million-in-race-toward-a-cancer-blood-test/#6b4cfccc4ac6.

Incisive Health (2014) *Saving lives, averting costs. An analysis of the financial implications of achieving earlier diagnosis of colorectal, lung and ovarian cancer.* London, UK: Incisive Health, September 2014. Available online: http://www.incisivehealth.com/uploads/Saving%20lives%20averting%20costs.pdf.

Iqbal, J.; Ginsburg, O.; Rochon, P.A.; Sun, P.; Narod, S.A. (2015) Differences in Breast Cancer Stage at Diagnosis and Cancer-Specific Survival by Race and Ethnicity in the United States. *JAMA: The Journal of the American Medical Association* **313**(2): 165-173. Available online: http://jamanetwork.com/journals/jama/fullarticle/2089353#88919843.

Laudicella, M.; Walsh, B.; Burns, E.; Smith, P.C. (2016) Cost of care for cancer patients in England: evidence from population-based patient-level data. *British Journal of Cancer* **114**(11): 1286-1292. Available online: https://www.ncbi.nlm.nih.gov/pmc/articles/PMC4891510/pdf/bjc201677a.pdf.

Lindahl, T. (1993) Instability and decay of the primary structure of DNA. *Nature* **362**(6422): 709-715.

Loeb, L.A.; Harris, C.C. (2008) Advances in chemical carcinogenesis: a historical review and perspective. *Cancer Research* **68**(17): 6863-6872. Available online: https://www.ncbi.nlm.nih.gov/pmc/articles/PMC2583449/pdf/nihms70190.pdf.

Mariotto, A.B.; Yabroff, K.R.; Shao, Y.; Feuer, E.J.; Brown, M.L. (2011) Projections of the Cost of Cancer Care in the United States: 2010-2020. *Journal of the National Cancer Institute* **103**(2): 117-128. Erratum in: *Journal of the National Cancer Institute* **103**(8): 699. Available online: https://www.ncbi.nlm.nih.gov/pmc/articles/PMC3107566/pdf/djq495.pdf.

Mehrotra, A.; Zaslavsky, A.M.; Ayanian, J.Z. (2007) *Archives of Internal Medicine* **167**(17): 1876-1883. Available online: http://jamanetwork.com/journals/jamainternalmedicine/fullarticle/486857.

Morris, C.R.; Ramirez, C.N.; Cook, S.N.; Parikh-Patel, A.; Kizer, K.W.; Bates, J.H.; Snipes, K.P. (2013) *Cancer Stage at Diagnosis, 2013.* Sacramento, CA: California Department of Public Health, California Cancer Registry, June 2013. Available online: www.ccrcal.org/pdf/Reports/CCR_Cancer_Stage_at_Diagnosis_2013.pdf.

Nasdaq GlobeNewswire (2017) GRAIL Closes Over $900 Million Initial Investment in Series B Financing to Develop Blood Tests to Detect Cancer Early. Available online: https://globenewswire.com/news-release/2017/03/01/929515/0/en/GRAIL-Closes-Over-900-Million-Initial-Investment-in-Series-B-Financing-to-Develop-Blood-Tests-to-Detect-Cancer-Early.html.

NCCN (2017) *NCCN Clinical Practice Guidelines in Oncology.* Fort Washington, PA: National Comprehensive Cancer Network. Available online: https://www.nccn.org/professionals/physician_gls/f_guidelines.asp.

NCI (2017a) *Surveillance, Epidemiology, and End Results Program: SEER Registry Groupings for Analyses.* Bethesda, MD: National Institutes of Health, National Cancer Institute. Available online: https://seer.cancer.gov/registries/terms.html.

NCI (2017b) *National Expenditures for Cancer Care*. Bethesda, MD: National Institutes of Health, National Cancer Institute. Available online: https://costprojections.cancer.gov/expenditures.html.

NCI (2017c) *Average Annual Costs of Care*. Bethesda, MD: National Institutes of Health, National Cancer Institute. Available online: https://costprojections.cancer.gov/annual.costs.html.

NCI (2017d) *Surveillance, Epidemiology, and End Results Program. Cancer Stat Facts: Cancer of Any Site.* Bethesda, MD: National Institutes of Health, National Cancer Institute. Available online: https://seer.cancer.gov/statfacts/html/all.html.

Ortman, J.M.; Velkoff, V.A.; Hogan, H. (2014) *An Aging Nation: The Older Population in the United States.* Current Population Reports, P25-1140. Washington, DC: U.S. Census Bureau, May 2014. Available online: https://www.census.gov/prod/2014pubs/p25-1140.pdf.

Parikh-Patel, A., Morris, C.R.; Martinsen, R., Kizer, K.W. (2015) *Disparities in Stage at Diagnosis, Survival, and Quality of Cancer Care in California by Source of Health Insurance.* Sacramento, CA: California Cancer Reporting and Epidemiologic Surveillance Program, Institute for Population Health Improvement, University of California Davis. Available online: https://www.ucdmc.ucdavis.edu/iphi/resources/1117737_cancerhi_100615.pdf.

Pyenson, B.S.; Fitch, K.V.; Pelizzari, P.M. (2016) *Cost drivers of cancer care: A retrospective analysis of Medicare and commercially insured population claim data 2004-2014.* Seattle, WA: Milliman, April 2016. Available online: http://www.milliman.com/uploadedFiles/insight/2016/trends-in-cancer-care.pdf.

Regalado A. (2017) Grail's $1 Billion Bet on the Perfect Cancer Test. *MIT Technology Review* (June 5, 2017). Available online: https://www.technologyreview.com/s/607944/grails-1-billion-bet-on-the-perfect-cancer-test/.

Schrag, D. (2015) *Cancer Care Spending in California: What Medicare Data Say.* Sacramento, CA: California HealthCare Foundation, August 2015. Available online: http://www.chcf.org/resources/download.aspx?id={3745E880-94E2-4216-89C8-F3E7ED3BABCC}.

Siegel, R.L.; Miller, K.D.; Jemal, A. (2017) Cancer statistics, 2017. *CA: A Cancer Journal for Clinicians* **67**(1): 7-30. Available online: http://onlinelibrary.wiley.com/doi/10.3322/caac.21387/full.

Styperek, A.; Kimball, A.B. (2012) Malignant Melanoma: The Implications of Cost for Stakeholder Innovation. *The American Journal of Pharmacy Benefits* **4**(2): 66-76. Available online: http://www.ajpb.com/journals/ajpb/2012/ajpb_marapr2012/malignant-melanoma-the-implications-of-cost-for-stakeholder-innovation.

Tomasetti, C., Li, L. and Vogelstein, B. (2017) Stem cell divisions, somatic mutations, cancer etiology, and cancer prevention. *Science* **355**(6331): 1330-1334.

**Tables**

| Stage at diagnosis | # of patients at diagnosis | 0-6 months postdiagnosis | 0-12 months postdiagnosis | 0-18 months postdiagnosis | 0-24 months postdiagnosis |
|---|---|---|---|---|---|
| 0 | 2300 | $48,477 | $60,637 | $67,450 | $71,909 |
| I/II | 4425 | $61,621 | $82,121 | $91,109 | $97,066 |
| III | 1134 | $84,481 | $129,387 | $147,470 | $159,442 |
| IV | 501 | $89,463 | $134,682 | $162,086 | $182,655 |
| All | 8360 | $62,774 | $85,772 | $96,499 | $103,735 |

**Table 1.** Average per-patient allowed costs, by disease stage at diagnosis, for breast cancer patients studied in [Blumen *et al*, 2016].

| Stage at diagnosis | Total | Non-Hispanic White | Hispanic White | Black | Chinese | Japanese | South Asian | Other Asian | Other |
|---|---|---|---|---|---|---|---|---|---|
| I | 48.0 | 50.8 | 40.1 | 37.0 | 50.1 | 56.1 | 40.4 | 45.2 | 43.6 |
| II | 34.6 | 33.2 | 38.7 | 38.6 | 35.7 | 32.4 | 38.7 | 38.1 | 37.2 |
| III | 12.4 | 11.4 | 15.9 | 16.6 | 10.7 | 8.5 | 15.3 | 12.4 | 13.5 |
| IV | 5.0 | 4.6 | 5.3 | 7.8 | 3.5 | 3.0 | 5.6 | 4.3 | 5.7 |

**Table 2.** Breast cancer incidence rates (in %) by stage at diagnosis from 2004 to 2011 for women who were identified in the Surveillance, Epidemiology, and End Results (SEER) 18 registries database, as reported in [Iqbal *et al*, 2015]. Columns 3-10 correspond to the eight racial/ethnic groups identified therein. Column 2 corresponds to all racial/ethnic groups.

| Cancer | Total, # | Stage 0, # | Stage I, # | Stage II, # | Stage III, # | Stage IV, # | Stage ?, # | Stage 0, % | Stage I, % | Stage II, % | Stage III, % | Stage IV, % | Stage ?, % | Stage I, % | Stage II, % | Stage III, % | Stage IV, % |
|---|---|---|---|---|---|---|---|---|---|---|---|---|---|---|---|---|---|
| Breast | 141654 | 27344 | 51515 | 37083 | 13360 | 5415 | 6937 | 19.3 | 36.37 | 26.18 | 9.43 | 3.82 | 4.9 | 47.98 | 34.54 | 12.44 | 5.04 |
| Cervix | 7454 | NA | 3516 | 894 | 1402 | 942 | 700 | NA | 47.17 | 11.99 | 18.81 | 12.64 | 9.39 | 52.06 | 13.23 | 20.76 | 13.95 |
| Colon | 55378 | 4872 | 11342 | 13393 | 12000 | 9849 | 3922 | 8.8 | 20.48 | 24.18 | 21.67 | 17.79 | 7.08 | 24.35 | 28.74 | 25.76 | 21.15 |
| Rectum | 22468 | 2045 | 5132 | 3825 | 4631 | 3406 | 3429 | 9.1 | 22.84 | 17.02 | 20.61 | 15.16 | 15.26 | 30.2 | 22.5 | 27.25 | 20.04 |
| Esophagus | 6786 | 118 | 1098 | 959 | 1063 | 2154 | 1394 | 1.74 | 16.18 | 14.13 | 15.66 | 31.74 | 20.54 | 20.82 | 18.18 | 20.15 | 40.84 |
| Kidney | 23664 | 373 | 12100 | 2193 | 3084 | 4070 | 1844 | 1.58 | 51.13 | 9.27 | 13.03 | 17.2 | 7.79 | 56.42 | 10.23 | 14.38 | 18.98 |
| Larynx | 4803 | 402 | 1647 | 644 | 606 | 1080 | 424 | 8.37 | 34.29 | 13.41 | 12.62 | 22.49 | 8.83 | 41.41 | 16.2 | 15.24 | 27.16 |
| Liver | 15246 | NA | 3964 | 2126 | 2682 | 2433 | 4041 | NA | 26 | 13.94 | 17.59 | 15.96 | 26.51 | 35.38 | 18.97 | 23.94 | 21.72 |
| Lung | 86954 | 34 | 14847 | 3083 | 18639 | 37467 | 12884 | 0.04 | 17.07 | 3.55 | 21.44 | 43.09 | 14.82 | 20.05 | 4.17 | 25.18 | 50.61 |
| Melanoma | 59676 | 23920 | 22250 | 3990 | 1910 | 1355 | 6251 | 40.08 | 37.28 | 6.69 | 3.2 | 2.27 | 10.47 | 75.39 | 13.53 | 6.47 | 4.59 |
| Oral | 18434 | 445 | 3272 | 2074 | 2463 | 6415 | 3765 | 2.41 | 17.75 | 11.25 | 13.36 | 34.8 | 20.42 | 23 | 14.58 | 17.31 | 45.1 |
| Ovary | 14295 | NA | 3427 | 870 | 3984 | 2995 | 3019 | NA | 23.97 | 6.09 | 27.87 | 20.95 | 21.12 | 30.39 | 7.72 | 35.33 | 26.56 |
| Pancreas | 19545 | 77 | 1248 | 3995 | 1331 | 9054 | 3840 | 0.39 | 6.39 | 20.44 | 6.81 | 46.32 | 19.65 | 7.99 | 25.56 | 8.52 | 57.93 |
| Prostate | 109601 | NA | 134 | 84673 | 7283 | 7097 | 10414 | NA | 0.12 | 77.26 | 6.65 | 6.48 | 9.5 | 0.13 | 85.37 | 7.35 | 7.16 |
| Stomach | 13566 | 140 | 2855 | 1269 | 1319 | 5014 | 2969 | 1.03 | 21.05 | 9.35 | 9.72 | 36.96 | 21.89 | 27.31 | 12.13 | 12.61 | 47.95 |
| Testis | 4809 | 11 | 3249 | 454 | 717 | 0 | 378 | 0.23 | 67.56 | 9.44 | 14.91 | 0 | 7.86 | 73.51 | 10.27 | 16.22 | 0 |
| Thyroid | 17968 | NA | 11375 | 1466 | 2134 | 1890 | 1103 | NA | 63.31 | 8.16 | 11.88 | 10.52 | 6.14 | 67.45 | 8.69 | 12.66 | 11.21 |
| Bladder | 31628 | --- | 22875 | 3434 | 1401 | 2331 | 1587 | --- | 72.33 | 10.86 | 4.43 | 7.37 | 5.02 | 76.15 | 11.43 | 4.66 | 7.76 |
| Uterus | 21710 | 242 | 13366 | 1537 | 2546 | 1447 | 2572 | 1.11 | 61.57 | 7.08 | 11.73 | 6.67 | 11.85 | 70.74 | 8.13 | 13.48 | 7.66 |

**Table 3.** Incidence numbers (column 2 = total, columns 3-7 = stages 0-IV, column 8 = stage unknown or "?", for each row columns 3 through 8 add up to column 2), incidence rates in % for stages 0-IV and stage unknown (columns 9-13 = stages 0-IV, column 14 = stage unknown, for each row columns 9 through 14 add up to 100% up to rounding to 2 digits), and incidence rates for stages I-IV only with stage 0 and stage unknown specifically excluded (columns 15-18 = stages I-IV, for each row columns 15 through 18 add up to 100% up to rounding to 2 digits). The incidence numbers are based on the data for California adults aged 20 and older diagnosed with the cancers listed in column 1 during 2005-2009 as reported in [Morris *et al*, 2013]. Some cancer sites in column 1 are abbreviated as follows: Breast = female breast, Cervix = cervix uteri, Rectum = rectum and rectosigmoid junction, Kidney = kidney and renal pelvis, liver = liver and intrahepatic bile ducts, Lung = lung and bronchus, Melanoma = melanoma of the skin, Oral = oral cavity and pharynx, Bladder = urinary bladder, Uterine = uterine corpus. For Bladder, stage I includes stage 0 (stage I = stage 0/I).

| Group | Total, # | Stage 0, # | Stage I, # | Stage II, # | Stage III, # | Stage IV, # | Stage ?, # | Stage 0, % | Stage I, % | Stage II, % | Stage III, % | Stage IV, % | Stage ?, % | Stage I, % | Stage II, % | Stage III, % | Stage IV, % |
|---|---|---|---|---|---|---|---|---|---|---|---|---|---|---|---|---|---|
| NHW | 91951 | 17315 | 36067 | 23185 | 7899 | 3304 | 4181 | 18.83 | 39.22 | 25.21 | 8.59 | 3.59 | 4.55 | 51.19 | 32.91 | 11.21 | 4.69 |
| Black | 8804 | 1634 | 2548 | 2533 | 1073 | 549 | 467 | 18.56 | 28.94 | 28.77 | 12.19 | 6.24 | 5.3 | 38.01 | 37.79 | 16.01 | 8.19 |
| Hispanic | 22856 | 4155 | 6887 | 6663 | 2915 | 988 | 1248 | 18.18 | 30.13 | 29.15 | 12.75 | 4.32 | 5.46 | 39.46 | 38.18 | 16.7 | 5.66 |
| Asian/PI | 16251 | 3818 | 5515 | 4353 | 1359 | 525 | 681 | 23.49 | 33.94 | 26.79 | 8.36 | 3.23 | 4.19 | 46.93 | 37.04 | 11.56 | 4.47 |
| Age 20-44 | 16560 | 3071 | 4245 | 5466 | 2445 | 627 | 706 | 18.54 | 25.63 | 33.01 | 14.76 | 3.79 | 4.26 | 33.21 | 42.76 | 19.13 | 4.9 |
| Age 45-64 | 69521 | 15088 | 24218 | 18211 | 6812 | 2590 | 2602 | 21.7 | 34.84 | 26.19 | 9.8 | 3.73 | 3.74 | 46.72 | 35.14 | 13.14 | 5 |
| Age 65+ | 55573 | 9185 | 23052 | 13406 | 4103 | 2198 | 3629 | 16.53 | 41.48 | 24.12 | 7.38 | 3.96 | 6.53 | 53.91 | 31.35 | 9.6 | 5.14 |

**Table 4.** Incidence numbers (column 2 = total, columns 3-7 = stages 0-IV, column 8 = stage unknown or "?", for each row columns 3 through 8 add up to column 2), incidence rates in % for stages 0-IV and stage unknown (columns 9-13 = stages 0-IV, column 14 = stage unknown, for each row columns 9 through 14 add up to 100% up to rounding to 2 digits), and incidence rates for stages I-IV only with stage 0 and stage unknown specifically excluded (columns 15-18 = stages I-IV, for each row columns 15 through 18 add up to 100% up to rounding to 2 digits). The incidence numbers are based on the data for California women aged 20 and older diagnosed with Breast Cancer during 2005-2009 as reported in [Morris *et al*, 2013]. The demographic groups are abbreviated as follows: NHW = Non-Hispanic White, PI = Pacific Islander.

| Cancer | Total, # | Stage 0, # | Stage I, # | Stage II, # | Stage III, # | Stage IV, # | Stage ?, # | Stage 0, % | Stage I, % | Stage II, % | Stage III, % | Stage IV, % | Stage ?, % | Stage I, % | Stage II, % | Stage III, % | Stage IV, % |
|---|---|---|---|---|---|---|---|---|---|---|---|---|---|---|---|---|---|
| Breast | 260590 | 49540 | 96601 | 67953 | 24317 | 10382 | 11797 | 19.01 | 37.07 | 26.08 | 9.33 | 3.98 | 4.53 | 48.48 | 34.1 | 12.2 | 5.21 |
| Colon | 97947 | 8235 | 20085 | 23615 | 21597 | 17961 | 6454 | 8.41 | 20.51 | 24.11 | 22.05 | 18.34 | 6.59 | 24.12 | 28.36 | 25.94 | 21.57 |
| Rectum | 30334 | 2814 | 7536 | 4794 | 6081 | 4242 | 4867 | 9.28 | 24.84 | 15.8 | 20.05 | 13.98 | 16.04 | 33.27 | 21.16 | 26.84 | 18.73 |
| Lung | 155820 | 142 | 27008 | 7381 | 31097 | 69944 | 20248 | 0.09 | 17.33 | 4.74 | 19.96 | 44.89 | 12.99 | 19.94 | 5.45 | 22.96 | 51.65 |
| Prostate | 198043 | NA | 16113 | 135698 | 14194 | 13359 | 18679 | NA | 8.14 | 68.52 | 7.17 | 6.75 | 9.43 | 8.98 | 75.66 | 7.91 | 7.45 |

**Table 5.** Incidence numbers and rates with the same conventions for columns 2-18 as in Table 4. The incidence numbers are based on the data for persons diagnosed with breast, colon, rectal, lung, and prostate cancer in California during 2004-2012 as reported in [Parikh-Patel *et al*, 2015]. Some cancer sites in column 1 are abbreviated as follows: Breast = female breast, Rectum = rectum and rectosigmoid junction, Lung = lung and bronchus.

| Cancer | Stage I, $ | Stage II, $ | Stage III, $ | Stage IV, $ | Stage I, % | Stage II, % | Stage III, % | Stage IV, % |
|---|---|---|---|---|---|---|---|---|
| Breast | 29,377 | 40,989 | 57,155 | 67,038 | 52 | 32 | 10 | 6 |
| Prostate | --- | 26,505 | 30,541 | 44,591 | --- | 84 | 8 | 8 |
| Lung | 60,038 | 73,509 | 84,726 | 93,166 | 22 | 4 | 26 | 48 |
| Colorectal | 49,189 | 66,613 | 83,980 | 108,599 | 25 | 29 | 26 | 20 |

**Table 6.** Medicare spending per patient in the first year after diagnosis (columns 2-5) and fractions of patients (columns 6-9) by stage at diagnosis based on California beneficiaries diagnosed in 2007-2011 and followed through 2012, as reported in [Schrag, 2015]. Spending estimates are based on Medicare fee-for-service patients only and are adjusted for inflation to 2013 dollars. For prostate cancer stages I and II are combined due to small numbers for stage I.

| Cancer | Stage I, $ | Stage II, $ | Stage III, $ | Stage IV, $ | Stage I, % | Stage II, % | Stage III, % | Stage IV, % |
|---|---|---|---|---|---|---|---|---|
| Breast | 64,889 | 70,931 | 71,555 | 70,057 | 27 | 32 | 19 | 22 |
| Prostate | --- | 66,160 | 82,621 | 71,704 | --- | 66 | 5 | 29 |
| Lung | 82,621 | 78,091 | 74,186 | 65,907 | 13 | 3 | 27 | 57 |
| Colorectal | 83,135 | 84,098 | 86,789 | 79,552 | 14 | 21 | 26 | 39 |

**Table 7.** Medicare spending per patient in the last year of life (columns 2-5) and fractions of patients (columns 6-9) by stage at diagnosis based on California beneficiaries who were diagnosed in 2007-2011 and died in 2007-2012, as reported in [Schrag, 2015]. Estimates include the full year of Medicare spending prior to and including the month of death, irrespective of when the patient was diagnosed. Spending estimates are based on Medicare fee-for-service patients only and are adjusted for inflation to 2013 dollars. For prostate cancer stages I and II are combined due to small numbers for stage I.

| Cancer | Average Costs, $ | Average Cost-Savings, $ | Average Cost-Savings, % |
|---|---|---|---|
| Breast | 38130 | 4330 | 11.35 |
| Prostate | 28275 | 1770 | 6.26 |
| Lung | 82897 | 20787 | 25.08 |
| Colorectal | 75170 | 16623 | 22.11 |

**Table 8.** Estimated average costs (second column), absolute cost-savings (third column) and relative cost-savings (in %, fourth column). These estimates are based on the data in Table 6.

| Stage | Average Costs, $ | Incidence rate, % |
|---|---|---|
| 0 | 984 | NA |
| I | 4259 | 52.1 |
| II | 12566 | 32.6 |
| III | 39761 | 9.7 |
| IV | 42303 | 5.6 |
| IV (recurrent) | 39281 | NA |

**Table 9.** Average costs in the first year after diagnosis (second column) and incidence rates by stage at diagnosis (third column) for malignant melanoma based on 2008 data as reported in [Styperek and Kimball, 2012]. The incidence rates are given for stages I-IV and add up to 100%.

| Cancer | $C_{12-mo}$ $ | $S_{12-mo}$ $ | $E_{12-mo}$ % | $X_{Early}$ $ | $X_{Late}$ $ | $G$ | $R_{Late}$ % | $F$ |
|---|---|---|---|---|---|---|---|---|
| Breast* | 90610 | 8489 | 9.37 | 82121 | 130909 | 0.5941 | 17.4 | 0.1034 |
| Breast | 38130 | 4330 | 11.35 | 33801 | 60861 | 0.8006 | 16 | 0.1281 |
| Prostate | 28275 | 1770 | 6.26 | 26505 | 37566 | 0.4173 | 16 | 0.0668 |
| Lung | 82897 | 20787 | 25.08 | 62110 | 90201 | 0.4523 | 74 | 0.3347 |
| Colorectal | 75170 | 16623 | 22.11 | 58546 | 94684 | 0.6172 | 46 | 0.2839 |
| Melanoma | 12541 | 5085 | 40.55 | 7456 | 40691 | 4.4573 | 15.3 | 0.682 |

**Table 10.** The quantities in columns 2-9 are defined in Subsection 6.1. Row 2 corresponds to the commercial claims data of [Blumen *et al*, 2016]. Rows 3-6 correspond to the Medicare data of [Schrag, 2015]. Row 7 corresponds to the data reported in [Styperek and Kimball, 2012].

| Cancer | Estimated National Spending in 2017, $M | SD, % | Estimated New Cases in 2017, # | Estimated Per-new-incidence Spending in 2017, $ | $G$ | $R_{Late}$ % | $E_{12-mo}$ % | Estimated National Cost-savings, $M |
|---|---|---|---|---|---|---|---|---|
| All Sites | 152901.1 | 8.49 | 1688780 | 90539 | --- | --- | --- | 25902 |
| Bladder | 4543.33 | 6.41 | 79030 | 57489 | *0.5202* | 12.42 | 6.07 | 276 |
| Brain | 5604.18 | 11.58 | 23800 | 235470 | *0.5232* | *39.94* | 17.28 | 968 |
| Breast | 19478.58 | 7.81 | 252710 | 77079 | **0.5941** | **17.4** | **9.37** | 1562 |
| Cervix | 1441.05 | 10.69 | 12820 | 112406 | *0.5232* | 34.71 | 15.37 | 221 |
| Colorectal | 15727.4 | 9.29 | 135430 | 116129 | **0.6172** | **46** | **22.11** | 3477 |
| Esophagus | 1857.85 | 15.09 | 16940 | 109672 | *0.5232* | 60.99 | 24.19 | 449 |
| Oral | 4101.55 | 9.34 | 49670 | 82576 | *0.5232* | 62.41 | 24.62 | 1010 |
| Kidney | 5487.4 | 13.12 | 63990 | 85754 | *0.5232* | 33.36 | 14.86 | 815 |
| Leukemia | 6772.22 | 8.72 | 62130 | 109001 | *0.5232* | *39.94* | 17.28 | 1170 |
| Lung | 13693.22 | 11.31 | 222500 | 61543 | **0.4523** | **74** | **25.08** | 3434 |
| Lymphoma | 15096.07 | 8.76 | 80500 | 187529 | *0.5232* | *39.94* | 17.28 | 2609 |
| Melanoma | 3308.32 | 10.28 | 87110 | 37979 | **4.4573** | **15.3** | **40.55** | 1342 |
| Ovary | 5338.73 | 11.32 | 22440 | 237911 | *0.5232* | 61.89 | 24.46 | 1306 |
| Pancreas | 3040.12 | 16.06 | 53670 | 56645 | *0.5232* | 66.45 | 25.8 | 784 |
| Prostate | 14873.72 | 5.47 | 161360 | 92177 | **0.4173** | **16** | **6.26** | 931 |
| Stomach | 2074.28 | 11.88 | 28000 | 74081 | *0.5232* | 60.56 | 24.06 | 499 |
| Uterus | 2947.42 | 9.08 | 61380 | 48019 | *0.5232* | 21.14 | 9.96 | 294 |
| Other | 27515.67 | 10.11 | 275300 | 99948 | *0.5232* | *39.94* | 17.28 | 4755 |

**Table 11.** Cancer site abbreviations in column 1 are the same as in Table 3. Column 2 = mean estimated spending based on [NCI, 2017b] (which uses the "head-and-neck" nomenclature for Oral = oral cavity and pharynx cancer), and column 3 = standard deviation (see Subsection 6.2). Column 4 = number of estimated new incidences as reported in [ACS, 2017] and [Siegel *et al*, 2017]. Column 5 = estimated per-new-incidence spending. The factors $G$ (column 6) and rate $R_{Late}$ (column 7) are taken from Table 10 for breast, colorectal, lung and prostate cancers and melanoma (bold font). For other cancer sites the factor $G$ is extrapolated from the values in Table 10 (italicized font) and rate $R_{Late}$ is taken from Table 3 (regular font) or extrapolated therefrom (italicized font). See Subsection 6.1. The relative cost-savings $E_{12-mo}$ (column 8) are obtained via Eq. (12) and Eq. (13), and cost-savings from early diagnosis (column 9) via Eq. (15), where the factor $H$ is set to 1 for all cancer sites except breast cancer, for which $H = 85.59\%$.